\documentclass[preprint,showpacs,superscriptaddress,aps]{revtex4-1}
\usepackage{amssymb}
\usepackage{amsmath,amssymb,graphicx}
\usepackage{graphicx}
\usepackage{dcolumn}
\usepackage{bm}
\def\be{\begin{equation}}
\def\ee{\end{equation}}
\def\bea{\begin{eqnarray}}
\def\eea{\end{eqnarray}}

\begin{document}

\title{Thermodynamic and cosmological parameters of early stages of the  Universe}

\author{Amr Abd Al-Rahman Youssef
} \email{amraay2003@gmail.com}
\affiliation{{\fontsize{10}{10}\selectfont{ Mathematics
Department, Faculty of Science, Al-Azhar University, Cairo,
Egypt.}}}

\author{Gaber Faisel}
\email{gaberfaisel@sdu.edu.tr}

\affiliation{{\fontsize{10}{10}\selectfont{Department of Physics,
Faculty of Arts and Sciences, S\"uleyman Demirel University,
Isparta, Turkey 32260.}}}

\author{ Hakan Akyildirim}
\email{hakanakyildirim@sdu.edu.tr}
\affiliation{{\fontsize{10}{10}\selectfont{Department of Physics,
Faculty of Arts and Sciences, S\"uleyman Demirel University,
Isparta, Turkey 32260.}}}

\begin{center}

\begin{abstract}
The early Universe was characterized by the presence of heavy
particles that decoupled at different temperatures leading to
different phases of the Universe. This had a consequences on the
time evolution of the thermodynamic and the cosmological
parameters characterizing each phase of the early Universe. In
this study, we derive the analytic expressions of the equations
governing the time evolution of these parameters in the early eras
of the Universe namely, the radiation era, the quark-gluon plasma
era, the hadron era and the mixed era. The parameters under
concern include the energy density, the entropy density, the
temperature, the pressure in addition to Hubble parameter and the
scale factor. Having these expressions allows us to give
estimations of the times corresponding to the beginning and ending
of each era of the Universe as will be presented in this work.
\end{abstract}
\end{center}

\maketitle

\section{Introduction} \label{sec:intr}

It is widely accepted that the Universe is homogeneous and
isotropic \cite{wein}. As a consequence, the space-time can be
parametrized by the Friedmann-Lema\^{i}tre-Robertson-Walker (FLRW)
metric. Upon inserting the metric into the Einstein equations one
obtains the Friedmann equations \cite{wein}. These equations can
be used to get the following equation
\cite{Ornik:1987up,Florkowski:2010mc,guardo}:
\begin{equation}
\frac{\dot{a}}{a}=-\frac{d\varepsilon}{3  \left( \varepsilon + p
\right)} = \sqrt{\frac{8 \pi\, G\, \varepsilon}{3}} dt
\label{eqdens}
\end{equation}
where $a(t)$ is the scale factor. The above differential equation
enables us to find the temporal variation of the energy density
$\varepsilon$ once the pressure $p$ as a function of $\varepsilon$
is known.  It should be noted that the above equation can be
rewritten in terms of temperature $T$ with the help of equations
of state and hence one gets solution expressing the temporal
evolution of the temperature. This can be the case also for other
thermodynamic parameters such as pressure density and entropy
density $s$.

The scale factor as a cosmological parameter can be calculated
with the help of the energy density and pressure upon performing
the integration in Eq.(\ref{eqdens}). On the other hand, knowing
the the energy density, the Hubble parameter $H(t)$ can be
estimated using the following equation

\be H(t)= \sqrt{\frac{8 \pi G}{3} \varepsilon(t)}\label{Hubp}\ee
Based on the above discussion, the time evolution of the
thermodynamic and cosmological parameters depends on the knowledge
of the equations of state of the Universe which in turn depend on
the phase of the Universe. This can be understood as in each phase
of the Universe, the nature of the matter spreading in the
Universe and the distribution of the energy in the Universe are
different.

It is widely believed that our Universe underwent different
cosmological phases  started right after the big bang and their
time evolution leaded to  our current  Universe. The early phase
of the Universe  was dominated by radiation. During this phase the
Universe endured several phase transitions due to the cooling
process to a temperature $T \sim m_c $ where $m_c$ is the charm
quark mass. At the end of radiation dominated (RD) era, the Universe experienced another
phase transition namely, quark-gluon plasma (QGP) phase. This
phase was followed by another phase the so called hadron phase
when the temperature of the Universe was smaller than the critical
temperature $T_C$ which represents the QCD phase transition
temperature. In the mixed phase the Universe experienced the
co-existence of QGP and hadron phases for a certain time interval.
This mixed phase happens when the temperature of the Universe
cools down and get close to $T_C$. During this phase, the
temperature of the Universe was fixed at $T_C$.

QGP can be created through colliding ultra-relativistic heavy ions
in Colliders as in AGS, SPS, RHIC, LHC. On the theoretical hand
side, there are two basic approaches for studying the properties
of QCD. The first approach is based on lattice Gauge Theories in
which  a field theory is formulated and solved on a discrete
lattice of space-time points with the help of so powerful
computers. As a prediction of the lattice QCD, the nuclear matter
experiences a phase transition at a temperature $T_C$ in the range
$150 - 170 MeV$ and energy density, $ ~ 1 GeV/fm^3$. The second
approach adopts phenomenological models to avoid the requirement
of the intense numerical calculations needed for lattice QCD.
Examples of such models include the bag models with the MIT bag
model is the widely used one. With the recent experimental results
from heavy ion collisions and the advances in lattice QCD
calculations  our knowledge of the equation of state of the QGP
has been improved.

In the literature, previous studies, related to studying early
phases of the Universe, have been carried out where the main
interest was directed to the QGP phase
\cite{Yagi:2005yb,Fogaca:2009wf,Florkowski:2010mc,Begun:2010eh,guardo,Sanches:2014gfa,Castorina:2015ava,Husdal:2016haj,Moradpour:2016tuw,Castorina:2018whj,McDonough:2020tqq,Elmashad:2021whh}.

In this study we aim to derive the analytic solutions of the
Eq.(\ref{eqdens}). The equation can be casted into energy density
 or temperature or pressure differential equation and thus can be
solved to give the time evolution of the corresponding
thermodynamic parameter in the early eras of the Universe as we
will show in details in the following. Moreover, we give more
attention to show details of estimating the times at which
different phase transitions occur and give analytic expressions
for estimating these times. We will also derive the expressions
governing the time variation of some cosmological parameters in
all of the aforementioned phases of the Universe.

\section{The time evolution of the
thermodynamic parameters in the early Universe}

In this section we investigate the thermodynamics and cosmological
parameters in early eras of the Universe. Our aim is to derive
analytic expressions for these parameters in each era. These
expressions can be used to study the time evolution of the
aforementioned parameters which will be presented in section
\ref{numa}.

The early Universe is thought to be characterized by very high
temperatures.  Consequently, massive particles were pair produced,
and contributed to the thermal bath. Moreover, particle masses can
be neglected providing that $m<< T$ where $m$ and $T$ denote the
mass of the particle and temperature of the Universe respectively.
In the Standard Model (SM) the heaviest particle is the top quark
with mass  $ m_t \simeq\,170 \, GeV$. Searches at the large Hadron
Collider (LHC) for particles with heavier masses than the top
quark mass exclude particles with masses close to TeV predicted in
many theories beyond the SM. Thus, our knowledge about phase
transitions occurred in the temperature interval $ T> m_t $
remains model dependent and is uncertain in the same time.
Consequently,  we adopt the SM as the framework in which the
evaluation of the degrees of freedom of particles and bosons,
required in this study, are carried out. In the SM, a chemical
potential is often associated with baryon number. Due to the fact
that the ratio of the net baryon density to the photon density is
so tiny, one can neglect that chemical potential when estimating
thermodynamic quantities such as the energy density $\varepsilon$,
the pressure density $p$ and the entropy density $s$. In the
following, we will present the equations of state that relate
these parameters with temperature and derive the equations
governing their time evolution in the early eras of the Universe.

\subsection{Radiation era}

The early epoch of the Universe was characterized by temperatures
satisfying the relation $ T> m_c $. In this epoch, the Universe
was dominated by radiation and endured several phase transitions
as a consequence of the cooling process to a temperature $T \sim
m_c $ where $m_c$ is the charm quark mass. The equations of state
in this era can be approximated as

\be \varepsilon_{RD} = N_{RD}\frac{\pi^2}{30}\, T^4 ,\,\,
\,\,\,\,\,\,\,\,\,\,\,\,\,\,\,\,\,\,\,\,\,\,\,\,\,\,\, p_{RD} =
N_{RD}\frac{\pi^2}{90}\, T^4, \,\,\,\,\,\,\,\,\,\,\,\,\,\, \,\,
\,\,\,\, \,\,\, s_{RD}=
\frac{(\varepsilon_{RD}+p_{RD})}{T}\,\,\,\,if\,\, m_c < T,
\label{edpppp} \ee here $N_{RD}$ stands for the effective number
of degrees of freedom at temperature $T$ and can be determined
from the relation \be N_{RD}= \sum_{B} g_{B}+\frac{7}{8} \sum_{F}
g_{F}\label{RDN}\ee where $g_{B}$ ($g_{F}$)  denotes number of
degrees of freedom for a boson $B$ (a fermion $F$) and the sum
runs over all boson and fermion states with masses satisfying $m<<
T$. Clearly, $N_{RD}$ is model dependent. The factor of $7/8$ in
the expression of $N_{RD}$ accounts for the difference between the
Bose and Fermi integrals.

 At high temperatures much bigger than the top quark mass $m_t$,
all the the SM particles were present. Thus, we have 28 bosonic
degrees of freedom and 90 fermionic degrees of freedom. The number
$28$ is the sum of degrees of freedom of the photons $\gamma$, the
charged gauge bosons $W^\pm$ , the neutral gauge boson $Z$, the
gluons $g$ and the Higgs boson $H$ where
$g_\gamma=2,g_{W^-}=g_{W^+}=g_Z =3,g_g=16$\, and $g_H =1$. On the
other hand, the number $90$ is the sum of degrees of freedom of
all fermions in the SM.  After substituting in Eq.(\ref{RDN}) we
find that $N_{RD}=28+\frac{7}{8}\times 90 =427/4=106.75 $

 We study the energy density in the time interval starting from $t_0=0$
 that corresponds to $T \sim \infty$
till time $t_8$  corresponding to $T_8 =m_c\sim 1 \,GeV$. First in
the mentioned time interval we have $p=c^2_s\,
\varepsilon_{RD}(t)$. Using this in Eq.(\ref{eqdens}) and setting
$c^2_s=1/3$ we get the following differential equation

\begin{equation}
\frac{d\varepsilon_{RD}}{\varepsilon_{RD} \sqrt{\varepsilon_{RD}}
} =- \sqrt{\frac{4\times 32 \pi G}{3}}\, dt \label{eqdens00}
\end{equation}
after integration we get

\be
\varepsilon_{RD}(t)=\frac{4\varepsilon_{RD}(t_i)}{\big[2+2\sqrt{\frac{32
\pi G}{3}\varepsilon_{RD}(t_i)}\, (t-t_i)\big]^2}\label{epsG0} \ee
where $t_i$ represents the initial time. The result agrees with
the the corresponding one given in Eq.(9) in Ref.
\cite{Castorina:2015ava}. For $i\to 0$ we find that
$\frac{1}{\sqrt{\varepsilon_{RD}(t_0)}}\to 0$, as $T(t_0)\to
\infty$, and hence we get \be \varepsilon_{RD}(t)= \frac{3}{ 32
\pi G\, t^2} \ee The preceding equation gives the expression of
the energy density in the time interval starting from $t=0$ to $t$
and for the intervals starting from $t_i\neq 0$ to $t_{i+1}$, the
energy density can be evaluated using Eq.(\ref{epsG0}). In the
temperature range starting at $T \sim \infty$ which corresponds to
the beginning of the Universe and ending at $T=m_c\sim 1 \,GeV$
several phase transitions occur due to the decoupling of heavy
particles and massive bosons. In particular, due to the decoupling
of top quark, Higgs boson, massive $Z$ and $W^\pm$ bosons, $b$
quark and $\tau$ lepton. This in turns will affect the value of
$N_{RD}$. In order to estimate the time at which these phase
transitions occur at temperature $T_i$, we can use the result of
the integration of Eq.(\ref{eqdens00}) and solve for $t$. Thus, we
find that \be t_{i+1}= t_i-\sqrt{\frac{3}{32 \pi G}}\,
\bigg(\frac{1}{\sqrt{\varepsilon_{RD}(t_i)}}-\frac{1}{\sqrt{\varepsilon_{RD}(t_{i+1})}}\bigg)\label{tall}\ee
It should be noted that in the time interval starting from $t_0=0$
till $t$ Eq.(\ref{tall}) reduces to \be t=  \sqrt{\frac{45}{ 16
\pi^3 G\, N_{RD}}}\,\frac{1}{T^2} \label{tedensRD}\ee in agreement
with Refs.\cite{Husdal:2016haj,ParticleDataGroup:2020ssz} after
setting $\hbar=c=k_{B}=1$. This relation can be used to estimate
the time at which all non standard particles, heavy particles
predicted in some classes of new physics beyond standard model,
decoupled. Since ongoing search at colliders has not observed such
particles up to $TeV$ energy scale, we can start our estimation of
times in the radiation era at $T=1\, TeV$ where only standard
model particles exist. In Table\ref{tim}, we present the numerical
estimation of the times at which standard model particles
decoupled and the corresponding energy densities.
\begin{table}
\begin{center}
\begin{tabular}{|c|c|c|c|c|}
  \hline
  i & $T_{i+1} (GeV)$ & $4\,N_{RD}$ & $t_{i+1} (s)$ & $ \varepsilon_{RD} (GeV/fm^3) $\\
  \hline
  0 & $ \,1000$ & 427 & $ 2.32\times 10^{-13} $ & $ 3.5\times 10^{13}$  \\
  1 & $ m_t$ & 385 & $ 8.22\times 10^{-12} $ & $ 3.7\times 10^{12}$  \\
 2 & $ m_H$  &381  &  $ 1.57\times 10^{-11} $ & $ 1.0\times 10^{12}$ \\
   3 & $ m_{Z^0}$  &369  & $ 2.96\times 10^{-11} $ & $  2.7\times 10^{11}$ \\
  4 & $  m_W^\pm$ & 345  & $ 3.97\times 10^{-11} $ & $ 1.5\times 10^{11}$ \\
  5 & $ m_b$ & 303 &   $  1.56\times 10^{-8} $ & $ 1.0\times 10^6$\\
  6 & $ m_\tau$ & 289  & $ 9.00\times 10^{-8} $ & $ 3.0\times 10^4$ \\
 7 & $ m_c$ & 247  & $ 1.85\times 10^{-7} $ & $ 7.3 \times 10^3$ \\
  \hline
\end{tabular}
\end{center}
\caption{ Time in seconds corresponding to decoupling of heavy
particles and massive weak gauge bosons.}\label{tim}
\end{table}

The time dependence of the temperature in the $RD$ phase of the
Universe can be obtained by substituting the definitions of
$\varepsilon_{RD}$ and $p_{RD}$ given in Eq.(\ref{edpppp}) into
Eq.(\ref{eqdens}) and hence, we obtain below simple differential
equation:
\begin{equation}
    T^{-3}\, dT = -\sqrt{\frac{4\pi^3 G N_{RD}}{45}}\, dt
    \label{eqTempDiffRD}
\end{equation}
we find that the solution of the above differential equation can
be expressed as
\begin{equation}
    T_{RD}(t) =
    \frac{T_{RD}(t_i)}{\left[1 + \sqrt{\dfrac{16 \pi^3 G N_{RD}T^4_{RD}(t_i)}{45}} (t -
    t_i)\right]^{1/2}}
        \label{eqTempRD00}
\end{equation}
where $T_{RD}(t_i)$ represents the initial temperature at the
start time $t=t_i$ of the time interval. At $i=0$ we have $
T_{RD}(t_0=0)=\infty $ and thus we can write

\begin{equation}
    T_{RD}(t) = \left[
\sqrt{\dfrac{16 \pi^3 G N_{RD}}{45}} \,\,t
    \right]^{-1/2}
    \label{eqTempRD1}
\end{equation}
this gives the evolution of temperature with time in the first
interval that ends at $T=1\, TeV$. For other intervals
corresponding to the times listed in Table\ref{tim}, we can use
the relation given in Eq.(\ref{eqTempRD00}) to estimate the
evolution of temperature with time.

The derivation of an analytic formula for the time variation of
the pressure in the radiation era is straightforward following
same steps as we did for the case of the energy density. The only
difference here is to replace $ \varepsilon_{RD}=3\, p_{RD}$ in
Eq.(\ref{eqdens}) and after performing the integration and setting
$\frac{1}{\sqrt{p_{RD}(t_0= \infty)}}\to 0$ we get \be p_{RD}(t)=
\frac{1}{ 32 \pi G\, t^2} \ee For any time interval starting at
$t=t_i$ we find that the pressure is given as \be
p_{RD}(t)=\frac{4 p_{RD}(t_i)}{3\big[2+2\sqrt{\frac{32 \pi
G}{3}\varepsilon_{RD}(t_i)}\, (t-t_i)\big]^2}\label{epsG} \ee

We turn now to derive the expression of the scale factor in the RD
era. Using the equations of state given in Eq.(\ref{edpppp}) and
Eq.(\ref{eqdens}) allows us to write
\begin{eqnarray}
{\frac{\dot{a}(t)}{a(t)}} &=& -\frac{\dot{\varepsilon}(t)}{3 \Big[
\varepsilon(t) + \frac{1}{3} \varepsilon(t) \Big]} = -\frac{1}{4}
\frac{\dot{\varepsilon}(t)}{\varepsilon(t)}\end{eqnarray} where we
have used  $p(t) = \frac{1}{3} \varepsilon (t) $. The previous
equation can be expressed as
\begin{eqnarray}
\frac{d}{dt}\ln \left[a(t) \right] &=& -\frac{1}{4}
\frac{d}{dt}\ln \left[ \varepsilon(t) \right]
\end{eqnarray}
Solution of such an equation yields

\begin{equation}
\frac{a(t)}{a(t_i)} = \Big[
\frac{\varepsilon(t_i)}{\varepsilon(t)} \Big]^{\frac{1}{4}}
=\big[1+\sqrt{\frac{32 \pi G}{3}\varepsilon_{RD}(t_i)}\,
(t-t_i)\big]^{\frac{1}{2}} \label{eqsf0}
\end{equation}
where, as before, $t_i$  stands for the  value of the time at the
beginning of the time interval in the RD era. It should be noted
that Eq.(\ref{eqsf0}) can be expressed in terms of the
temperatures or in terms of the times as

\begin{equation}
\frac{a(t)}{a(t_i)} = \Big[
\frac{\varepsilon(t_i)}{\varepsilon(t)} \Big]^{\frac{1}{4}}=
\frac{T (t_i)}{T (t)} = \big(\frac{t}{t_i}\big)^{\frac{1}{2}}
\label{eqsf11}
\end{equation}
The result obtained in the previous equation agrees with the
result obtained in Refs.\cite{Florkowski:2010mc,Sanches:2014gfa}.

\subsection{Quark Gluon Plasma era}

The QGP phase of the Universe existed when the temperature of the
Universe was in the range $ T_{C}< T < m_c $, where $T_C$ is the
critical temperature. In that phase, the Universe was in a state
filled with quark-gluon plasma contains lighter quarks in addition
to the photons, lighter charged leptons, neutrinos and
antineutrinos in thermal equilibrium. It should be noted that the
relativistic heavy ion collision experiments at both  RHIC and LHC
may access to the temperature range $ T_{C}< T < m_c $ and hence
they can shed light on the properties and nature of the plasma
formed in this range.

As it is known, quarks as colored particles are confined to each
others in bound hadronic states.  One of the most successful
phenomenological models for quark confinement is the so called MIT
bag model \cite{mit}. While in the MIT bag model  the
contributions that arise from the particles in the electroweak
sector were not taken into account, here in this work we follow
Refs.\cite{Florkowski:2010mc,Castorina:2015ava} and include their
effects on  the effective number of degrees of freedom. The
densities corresponding to this epoch of the early universe can be
approximated in a bag model $M_i$ as

\begin{eqnarray}
\varepsilon_{QGP} &=& N_{QGP}\frac{\pi^2}{30}\, T^4+
\mathcal{B},\,\, \,\,\,\,\,\,\,\, p_{QGP} = \frac{1}{3}
N_{QGP}\frac{\pi^2}{30}\, T^4- \mathcal{B},\,\,\,\,s_{QGP}= 4
N_{QGP}\frac{\pi^2}{90}\, T^3,\label{edQGP2}
\end{eqnarray}
here $\mathcal{B}$ is a bag constant parameter. It represents the
exerted external pressure on the bag surface.  This pressure
balances the internal pressure  in the absence of QGP and hence
ensures the stability of the bag. In Eq.(\ref{edQGP2}), $N_{QGP}$
stands for the effective number of degrees of freedom  and can be
determined from a relation similar to the one given in
Eq.(\ref{RDN}) where the summation in this case is carried out for
all the bosons and fermions present in the QGP.

The exact analytic solution of the energy density
$\varepsilon_{QGP}$, temperature and pressure densities  can be
obtained directly from solving Eq.(\ref{eqdens}). In the appendix,
we show the steps we follow to  derive  the desired solutions for
the three quantities. We find that, the analytic expressions can
be expressed as \bea \varepsilon_{QGP} (t)&=& \chi(t) +
\zeta(t)+\sqrt{\bigg(\chi (t)+\zeta (t)\bigg)^2-\zeta^2 (t)}
\label{edMT0}\eea The functions $\zeta(t)$ and $\chi(t)$ are given
in terms of $\eta(t)$ defined as

\be \eta(t) = \exp\bigg(4\,\sqrt{\frac{8 \pi \mathcal{B}\, G}{3}}\,\, t +
\xi \bigg)\ee with \bea \xi &=&
\ln\bigg(\frac{\sqrt{\varepsilon_{RD}
(t_8)}+\sqrt{\mathcal{B}}}{\sqrt{\varepsilon_{RD}
(t_8)}-\sqrt{\mathcal{B}}}\bigg)-4\,\sqrt{\frac{8 \pi
\mathcal{B}\, G}{3}}\,\, t_8 \label{xi}\eea where
$\varepsilon_{RD}(t_8)$ is the value of the energy density at the
time $t_8$. The explicit expressions of $\zeta(t)$ and $\chi(t)$
are given in Eq.(\ref{zeta}) in the appendix. It should be
remarked that, up to our knowledge, our analytic solution of the
energy density $\varepsilon_{QGP}$ given in Eq.(\ref{edMT0}) was
not pointed out in the literature before. Previous studies
reexpressed Eq.(\ref{eqdens}) in terms of temperature and solved
it analytically as in Ref.\cite{Yagi:2005yb} or numerically as in
Ref.\cite{Sanches:2014gfa} to obtain the temperature and
consequently used the equations of state to evaluate the energy
density.

The analytic solution of the temperature for the bag models of QGP
discussed above can be written as

\begin{equation} T_{QGP}(t)= \sqrt{\frac{2\, \mathcal{B}\kappa(t)}{\big(\frac{\pi^2}{30}N_{QGP}-\mathcal{B}\, \kappa^2 (t)
\big)}}\label{Tem}\end{equation} where the function $\kappa(t)$ is
given as
\begin{equation} \kappa(t) = b \exp\big[{-\frac{4}{3} \sqrt{6\pi \mathcal{B} G}\left( t-t_{8}\right)}\big]
\end{equation}
with \be b=
T^2_8\bigg(\frac{\mathcal{B}}{\frac{\pi^2}{30}N_{QGP}}+\sqrt{\frac{\mathcal{B}}{\frac{\pi^2}{30}N_{QGP}}
T^4_8+\frac{\mathcal{B}^2}{\frac{\pi^4}{900}N^2_{QGP}}}\bigg)^{-1}\ee
here $T_8= T_{RD}(t_8)$.

In Ref.\cite{Yagi:2005yb}, Eq.(\ref{eqdens}) was written in terms
of the critical temperature and solved analytically  to obtain the
time evolution of the temperature. Here, our result in
Eq.(\ref{Tem}) has no dependency on the critical temperature but
instead depends on the temperature at the beginning of the $QGP$
era which is the same one at the end of the radiation era.

Turning now to the pressure, Eq.(\ref{eqdens}) can be expressed in
terms of the pressure using equations of state and then can be
solved analytically to obtain the explicit dependency of the
pressure on the time as
\begin{equation}  p(t)= \frac{-3 \varrho_{8}^2 \mathcal{B} \exp\bigg[- 8\, \sqrt{\frac{8 \pi \mathcal{B} \, G}{3}} \big(
t-t_{8}\big)\bigg] -10 \varrho_{8} \mathcal{B} \exp\bigg[- 4\,
\sqrt{\frac{8 \pi \mathcal{B} \, G}{3}} \big( t-t_{8}\big)\bigg]
-3 \mathcal{B}} {3 \bigg(\varrho_{8} \exp\big[- 4\, \sqrt{\frac{8
\pi \mathcal{B} \, G}{3}} \big( t-t_{8}\big)\big]
+1\bigg)^2}\end{equation} where $\varrho_{8}=
\frac{\sqrt{\mathcal{B}}-\sqrt{3 p_{8} +4 \mathcal{B}
}}{\sqrt{\mathcal{B}}+\sqrt{3 p_{8} +4 \mathcal{B} }}$.

The scale factor in the $QGP$ era can be obtained with the help of
the equations of state listed in Eq.(\ref{edQGP2}) and
Eq.(\ref{eqdens}) and performing the integration. We find that
\begin{equation}
\frac{a(t)}{a(t_8)} = \Big[ \frac{\varepsilon_{QGP}(t_8) -
\mathcal{B}}{\varepsilon_{QGP}(t) - \mathcal{B}}
\Big]^{\frac{1}{4}}
=\frac{T_{QGP}(t_8)}{T_{QGP}(t)}=\frac{T_8}{T_{QGP}(t)}
\label{eqScalerFact_QGP}
\end{equation}
Thus, one can estimate the scale factor using either the
expression of the energy density or the expression of the
temperature in the $QGP$ era.

\subsection{Hadron era}

The critical temperature $T_C$ represents the QCD phase transition
temperature.  The  transition characterizes the
confinement-deconfinement transition between quarks and hadrons
where three quarks (anti-quarks) are confined together to form
baryon(anti-baryon) and quark anti-quark are confined together to
form meson. The formed heavy hadrons are not stable and thus
quickly decay to the lightest hadrons i.e. the pions. In the
hadron phase, the particle content includes
$e^\pm,\mu^\pm,\nu_{e,\mu,\tau},\bar \nu_{e,\mu,\tau} $ together
with photons and pions. The densities corresponding to the
massless pion gas are given as

\begin{eqnarray}
\varepsilon_{H} &=& N_{H}\frac{\pi^2}{30}\, T^4,\,\,
\,\,\,\,\,\,\,\,\,\,\,\,\,\,\,\,\,\,\,\,\, p_{H} =
N_{H}\frac{\pi^2}{90}\, T^4 \,\,\,\,\,\,\,\,\,\,\,\, \,\,\,\,
\,\,\,\,\, s_{H}=4 N_{H}\frac{\pi^2}{90}\, T^3\,\,\,\,\,\,\,\,\,
m_\pi < T < T_C\label{edH1}
\end{eqnarray}
where $m_\pi$ represents the pion mass and $N_{H}$ is the
effective number of degrees of freedom that can be calculated
using the relation

\be  N_{H}= N^2_f-1+ N_{EW}\label{QGP5}\ee where $N_f$ is the
number of flavors and $N_{EW}$ account for the contributions of
$e^\pm,\mu^\pm,\nu_{e,\mu,\tau},\bar \nu_{e,\mu,\tau} $ together
with photons.

Phase equilibrium is achieved when $p_{QGP}=p_{H}$ at $T=T_C$ and
thus we find, from Eq.(\ref{edQGP2}) and Eq.(\ref{edH1}), that

\be \mathcal{B} =\frac{\big(N_{QGP}-N_{H}\big)\,\pi^2}{90} T^4_C
 \ee

Time dependence of energy density in this era can be calculated
using Eq.(\ref{eqdens}) and Eq.(\ref{edH1}). It is clear from
Eq.(\ref{edH1}) that $p_H = \dfrac{1}{3}\varepsilon_H$, hence we
have the form
\begin{equation}
    -\frac{d\varepsilon_H}{\sqrt{\varepsilon_H} \left( \varepsilon_H + \frac{1}{3}\varepsilon_H \right)} = \sqrt{24\pi G}\, dt \label{eqdensHadron}
\end{equation}
Solving this equation we find the evolution of energy density as
follows
\begin{equation}
\varepsilon_{H}(t) = \left[ \dfrac{1}{\sqrt{\varepsilon_{H_0}}} +
\sqrt{\dfrac{32 \pi G}{3}} (t - t_{10}) \right]^{-2}
\end{equation}
Here $t_{10}$ is the time at which the hadron era started and
$\varepsilon_{H_0}$ is the initial energy density at the beginning
of hadronic era i.e. at $t_{10}$. On the other hand, we can obtain
the time dependence of temperature in hadron era by substituting
the definitions of $\varepsilon_{H}$ and $p_{H}$ given in
Eq.(\ref{edH1}) into Eq.(\ref{eqdens}). We obtain differential
equation similar to that one given in Eq.(\ref{eqTempDiffRD}) with
the only change $RD\to H$ and its solution yields
\begin{eqnarray}
T_{H}(t) &=& \left[ \dfrac{1}{{T^2_{H_0} }} + \sqrt{\dfrac{16
\pi^3 G N_H}{45}} (t - t_{10}) \right]^{-1/2}\nonumber\\
&=& \left[ \dfrac{1}{{T^2_C }} + \sqrt{\dfrac{16 \pi^3 G N_H}{45}}
(t - t_{10}) \right]^{-1/2} \label{eqTempHadronic0}
\end{eqnarray}
here $T_{H_0}$ is the initial temperature corresponding to
$t_{10}$ which is the same temperature $T_C$ at the end of the
mixed era. Following Ref.\cite{Yagi:2005yb}, we define the two
quantities $r$ and $\lambda$ as
\begin{eqnarray}
    r &=& \frac{N_{QGP}}{N_H} \nonumber\\
    \lambda &=& \sqrt{\frac{3}{8 \pi G \mathcal{B}}} \nonumber
    \label{eqTimeScaleLambda}
\end{eqnarray}
As stated in Ref.\cite{Yagi:2005yb}, the quantity  $r$ expresses a
number obtained in the pressure equilibrium condition, at $T=T_C$,
for the QGP and hadron phases while $\lambda$ is the time scale
for the QCD phase transition. In terms of $r$ and $\lambda$, we
can obtain a simple expression of the temperature as

\begin{eqnarray}
T_{H}(t) &=& T_C \bigg(1+\sqrt{\frac{12}{r-1}}
\frac{t-t_{10}}{\lambda}\bigg)^{-\frac{1}{2}}\label{eqTempHadronic}
\end{eqnarray}
It should be noted that, in obtaining the above result, we used
$\mathcal{B}=N_H(r-1)\frac{\pi^2}{90} T^4_C$. Using the expression
of $T_{H}(t)$ given in Eq.(\ref{eqTempHadronic}) we can obtain the
following expressions of the energy density and pressure in the
hadron era

\begin{eqnarray}
\varepsilon_{H}(t) &=&\frac{\pi^2}{30}\,N_H T^4_C
\bigg(1+\sqrt{\frac{12}{r-1}}
\frac{t-t_{10}}{\lambda}\bigg)^{-2}\nonumber\\
p_{H}(t) &=&\frac{\pi^2}{90}\,N_H T^4_C
\bigg(1+\sqrt{\frac{12}{r-1}}
\frac{t-t_{10}}{\lambda}\bigg)^{-2}\label{eqTempHadronic00}
\end{eqnarray}

We proceed now to find the equation governing the time evolution
of the scale factor in the hadron era.  The scale factor, then,
can be obtained with the help of the equations of state given in
Eq.(\ref{edH1}) and Eq.(\ref{eqdens}). Following the same steps
done in the RD era, substituting $p_H=\frac{1}{3} \varepsilon_H $
and performing the integration we obtain

\begin{equation}
\frac{a(t)}{a(t_{10})} =\frac{T_C}{T_{H}(t)}
\label{eqScalerFact_QGP0}
\end{equation}
using Eq.(\ref{eqTempHadronic}), we finally obtain
\begin{equation}
\frac{a(t)}{a(t_{10})} =\bigg(1+\sqrt{\frac{12}{r-1}}
\frac{t-t_{10}}{\lambda}\bigg)^{\frac{1}{2}}
\label{eqScalerFact_QGP}
\end{equation}
The above result agrees with the corresponding one shown in
Ref.\cite{Yagi:2005yb}.
\subsection{Mixed era\label{mixsub}}
In the transition from the QGP phase to the hadron phase, the
Universe experiences the co-existence of the both phases for a
certain time interval. This mixed phase happens when the
temperature of the Universe cools down and get close to  $T_C$.
During this time interval, the temperature of the system is fixed
at $T_C$.  This can be understood as the cooling of the Universe
due to its expansion is balanced by the release of the latent
heat. In the mixed phase, the energy density can be parameterized
as \cite{Yagi:2005yb}

\be \varepsilon (t) = \varepsilon_{H}(T_C) f(t) +
\varepsilon_{QGP} (T_C)\big (1-f(t)\big)\label{qgmix}\ee where
$f(t)$ takes the values $0(1)$ at the start (end) of the
co-existence. Regarding the pressure in the mixed era,  we find
that it can be parameterized in a similar way to the energy
density and thus can be written as \be p (t) =p_{H}(T_C) f(t) +
p_{QGP} (T_C)\big (1-f(t)\big)\label{qgmixp}\ee Using
Eqs.(\ref{eqdens}, \ref{qgmix}, \ref{qgmixp}), see the appendix
for detailed derivation, one obtains the following differential
equation
\begin{eqnarray}
\frac{df}{dt} &=& \frac{3}{\lambda} \left( \frac{r}{r-1} -
f\right)\, \sqrt{4 (1-f) + \frac{3}{r-1}}
\end{eqnarray}
The preceding equation has two analytic solutions that can be
expressed as
\begin{eqnarray} f_\pm(t) &=& 1 - \frac{1}{4 (r-1)}\left[
\tan^2 \left( \frac{3 (t-t_9)}{2\lambda \sqrt{r-1}} \pm
\tan^{-1}\sqrt{4r -1 } \right) -3 \right] \label{eqft0}
\end{eqnarray}
where $t_9$  stands for the beginning time of the mixed era. As we
will show in the following only $f_-(t)$ is the acceptable
solution and thus we take $f(t)=f_-(t)$. Having the expressions of
the energy density and pressure in the mixed era, we can now use
Eq.(\ref{eqdens}) to obtain an analytic expression of the scale
factor in the mixed era. We find that
\begin{equation}
\frac{\dot{a}(t)}{a(t)}= \frac{\dot{f}}{3
\big(\frac{r}{r-1}-f\big)}=\lambda^{-1}\sqrt{4(1-f)+\frac{3}{r-1}}
\label{fried2}
\end{equation}
in agreement with Ref.\cite{Yagi:2005yb}. Details about the
derivation of the above equation can be found in the appendix. The
equation has an analytic solution that can be obtained upon
performing the integration and can be expressed as
\cite{Yagi:2005yb}
\begin{equation}
\frac{a(t)}{a(t_9)} =(4r)^{\frac{1}{3}}\left[ \sin \left( \frac{3
(t-t_9)}{2\lambda \sqrt{r-1}} + \sin^{-1}\frac{1}{\sqrt{4 r}}
\right) \right]^{\frac{2}{3}} \label{fried3}
\end{equation}

\section{Numerical results and analysis \label{numa}}
We start our analysis by estimating the approximate times
corresponding to the ending of $QGP$, the mixed and the Hadron
phases of the early Universe i.e $t_9$, $t_{10}$ and $t_{11}$
respectively.

The value of $t_9$ can be determined from setting $T = T_C$ in
Eq.(\ref{Tem}) and solve for $t_9$. This leads to
\begin{equation} \kappa^2 (t_9)+\frac{2}{T^2_C} \kappa(t_9)-\frac{N_{QGP} \pi^2}{30\mathcal{B}}=0\label{Tg1}\end{equation}
The above equation has two solutions $ \kappa(t_9)= \kappa_1$ and
$\kappa(t_9)= \kappa_2$ where $  \kappa_{1,2}=-\frac{1}{T^2_C}
\pm\sqrt{ \frac{N_{QGP} \pi^2}{30\mathcal{B}} +\frac{1}{T^4_C}}$.
After solving for $t$ we get the two solutions \be
t_{9}=t_{8}-\frac{3}{4\sqrt{6\pi \mathcal{B}
G}}\ln\big(\frac{\kappa_{1,2}}{b}\big)\label{tqgpf}\ee After
setting $T_C= 170 MeV$, we find that $ \kappa_2$ is negative. This
leads to complex time and so this solution is not accepted. Thus
we are left with the other solution $\kappa_1\simeq
46.7\,GeV^{-2}$ that yields the time corresponds to the end of QGP
phase or the beginning of the mixed era as $ t_{9} \simeq 11.3\,
\mathrm{\mu s} $.

The time corresponding to the end of the mixed phase, $t_{10}$,
can be estimated from solving the equation $f_\pm(t_{10}) = 1$.
Using the expressions of $f_\pm(t)$ given in Eq.(\ref{eqft0}) and
upon setting $f_-(t_{10}) = 1$ we get
\begin{equation}
t_{10} - t_9 = \dfrac{2\lambda\sqrt{r-1}}{3} \left[
\tan^{-1}\sqrt(4r-1)-\tan^{-1}\sqrt{3} \right] \simeq 10.7\,
\mathrm{\mu s} \label{eqt9MixedEraEnd}
\end{equation}
Using the value $t_9\simeq 11.3\, \mathrm{\mu s}$ estimated before
we obtain $ t_{10} \simeq 22.0\, \mathrm{\mu s} $. It should be
noted that setting $f_+(t_{10}) = 1$ one leads to a value of
$t_{10}$ smaller than $ t_{9}$ which is not acceptable and thus
the function $f(t)=f_-(t) $ gives the correct behavior in the
mixed era in agreement with the choice of Ref.\cite{Yagi:2005yb}.

\begin{figure}[t]
\begin{center}
\includegraphics[width=7cm,height=7cm]{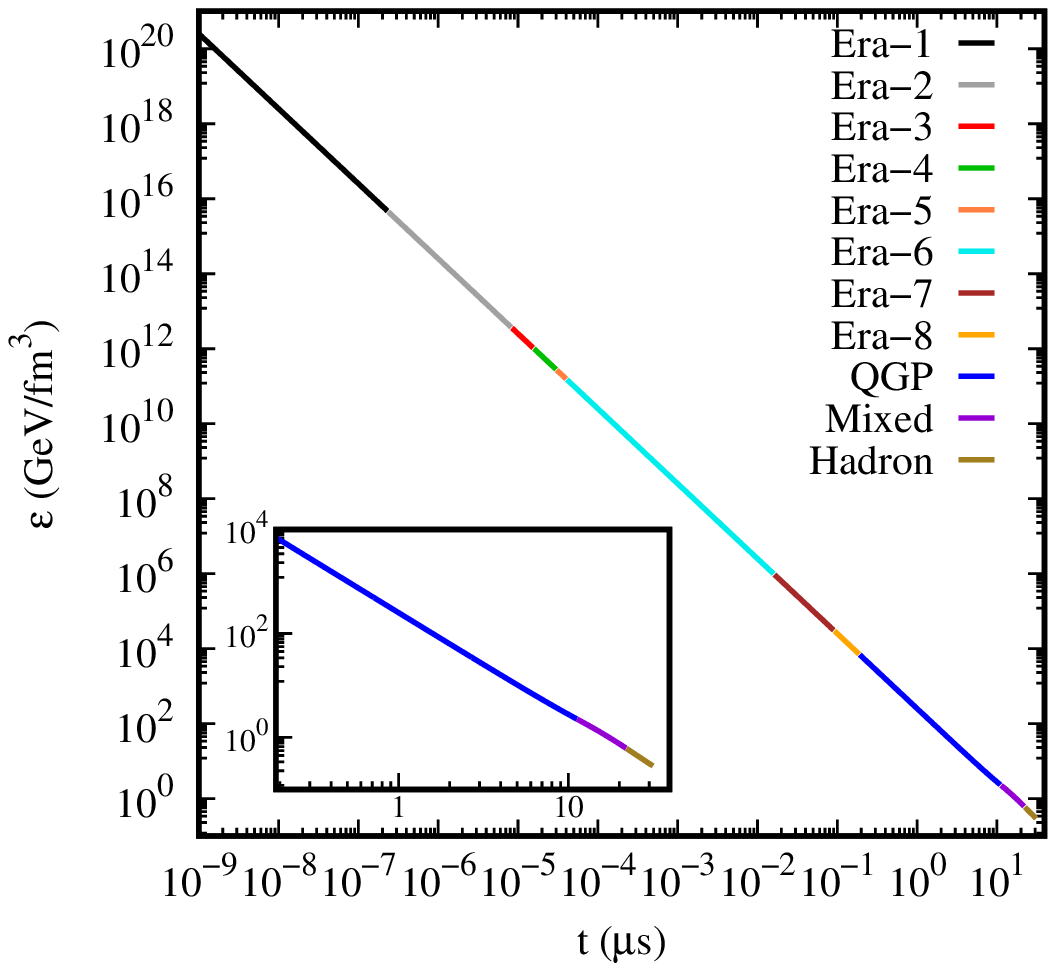}
\hspace{0.2cm}
\includegraphics[width=7cm,height=7cm]{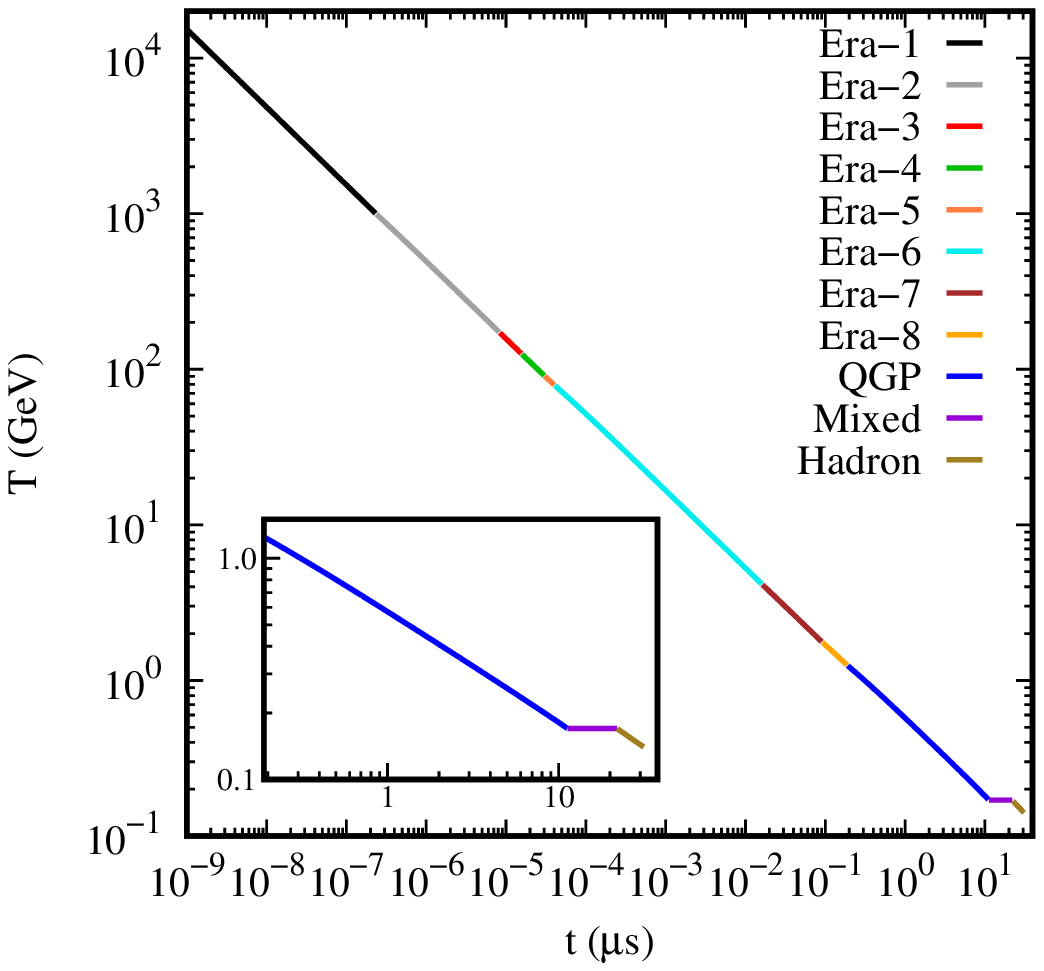}
\hspace{0.2cm}
\includegraphics[width=7cm,height=7cm]{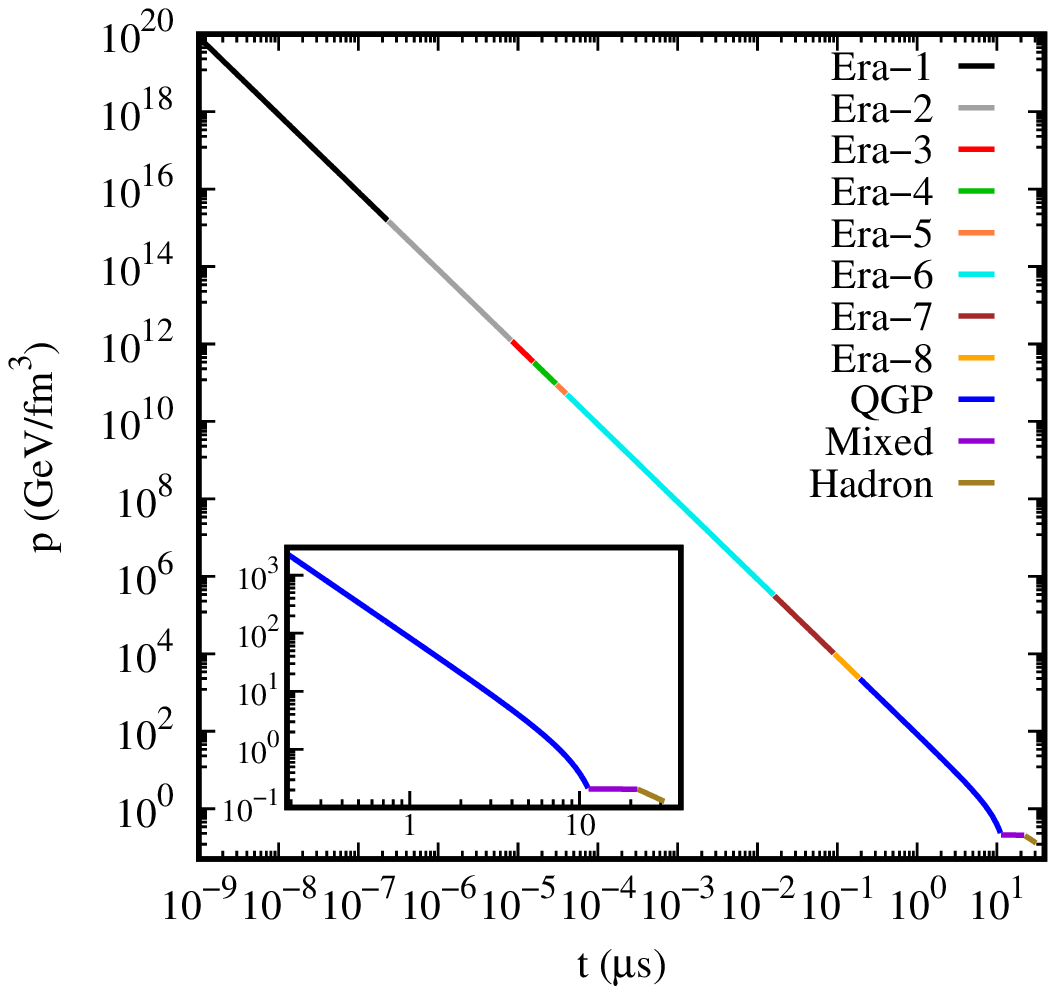}
\hspace{0.2cm}
\includegraphics[width=7cm,height=7cm]{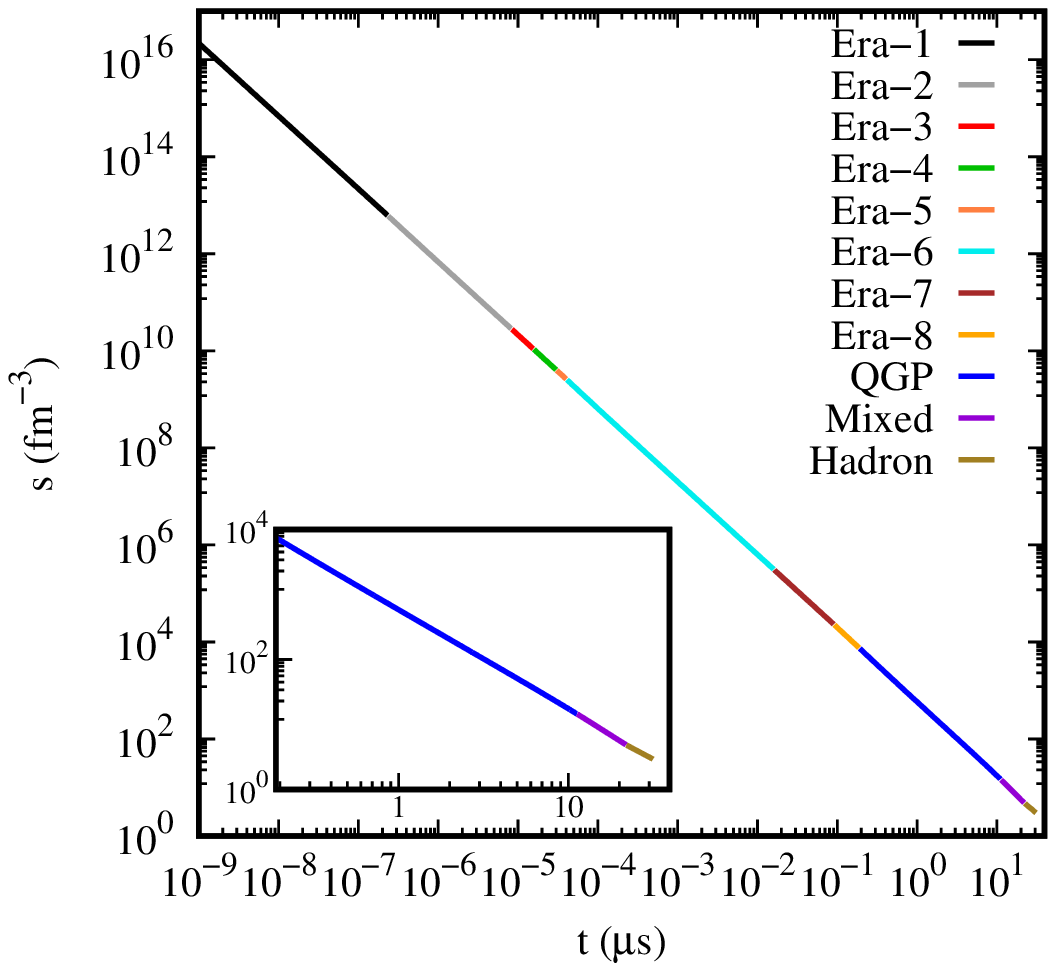}
\medskip
\caption{Time evolution of the energy density, pressure density,
entropy density and temperature in the four eras of the
Universe.\label{ETD}}
\end{center}
\end{figure}

The Hadron phase of the universe ends at a time $t_{11}$ which can
be calculated by setting $ T_{H}(t)= m_\pi$. The reason is
attributed to the remark that, for temperatures smaller than this
value most of the hadrons undergo either decays or annihilations
to final states containing lighter leptons or massless gauge
bosons.  Moreover, at these temperatures, only small amount of
protons and neutrons remain which can be deduced from the ratio
$\frac{n_B}{n_\gamma} = 6\times 10^{-10}$ where $n_B$ is the net
baryon number density and $n_\gamma$ is photon number density. For
$m_\pi=140\,GeV$, we find that the solution of the equation $
T_{H}(t)= m_\pi$ results in $t_{11}\simeq 31.5 \mathrm{\mu s}$.
Having determined all times corresponding to the beginning and the
ending of the radiation eras, given in Table \ref{tim}, $QGP$ era,
mixed and hadron eras, we are ready now to show our results for
the time variation of the thermodynamic and cosmological
parameters in all these eras.

\begin{figure}[t]
\begin{center}
\includegraphics[width=5cm,height=5cm]{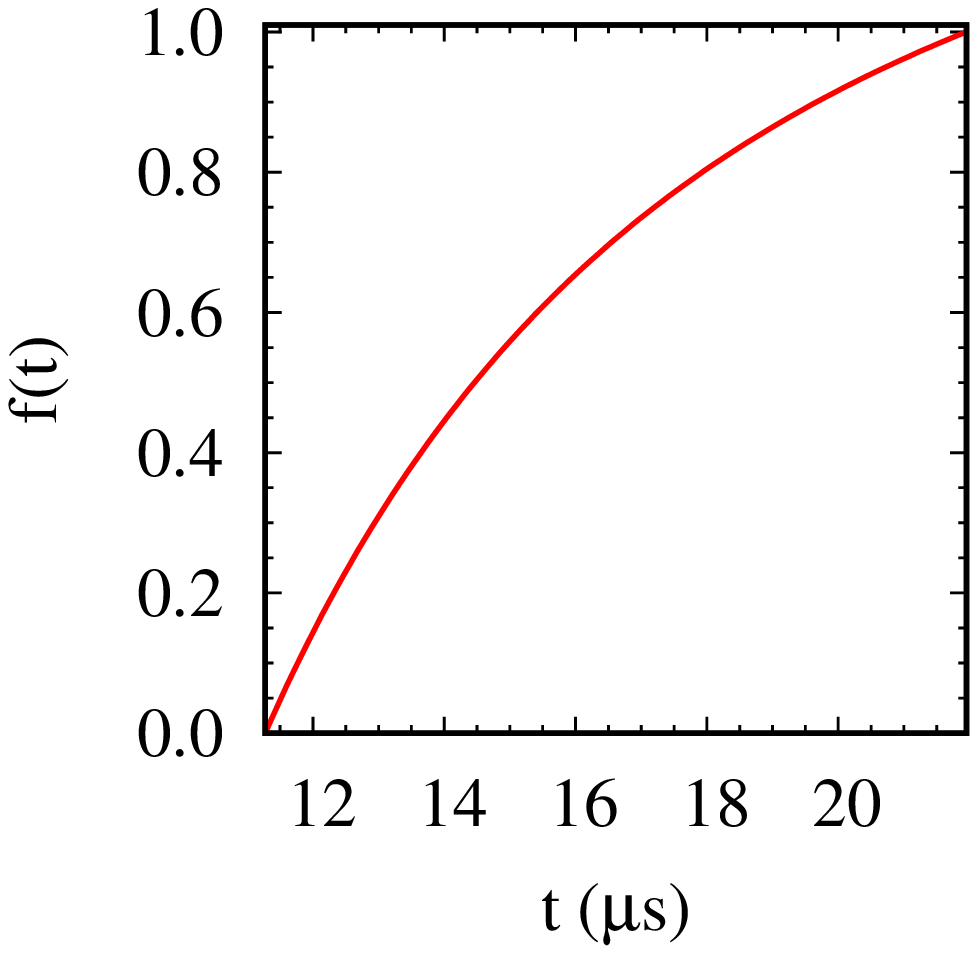}
\hspace{0.2cm}
\includegraphics[width=5cm,height=5cm]{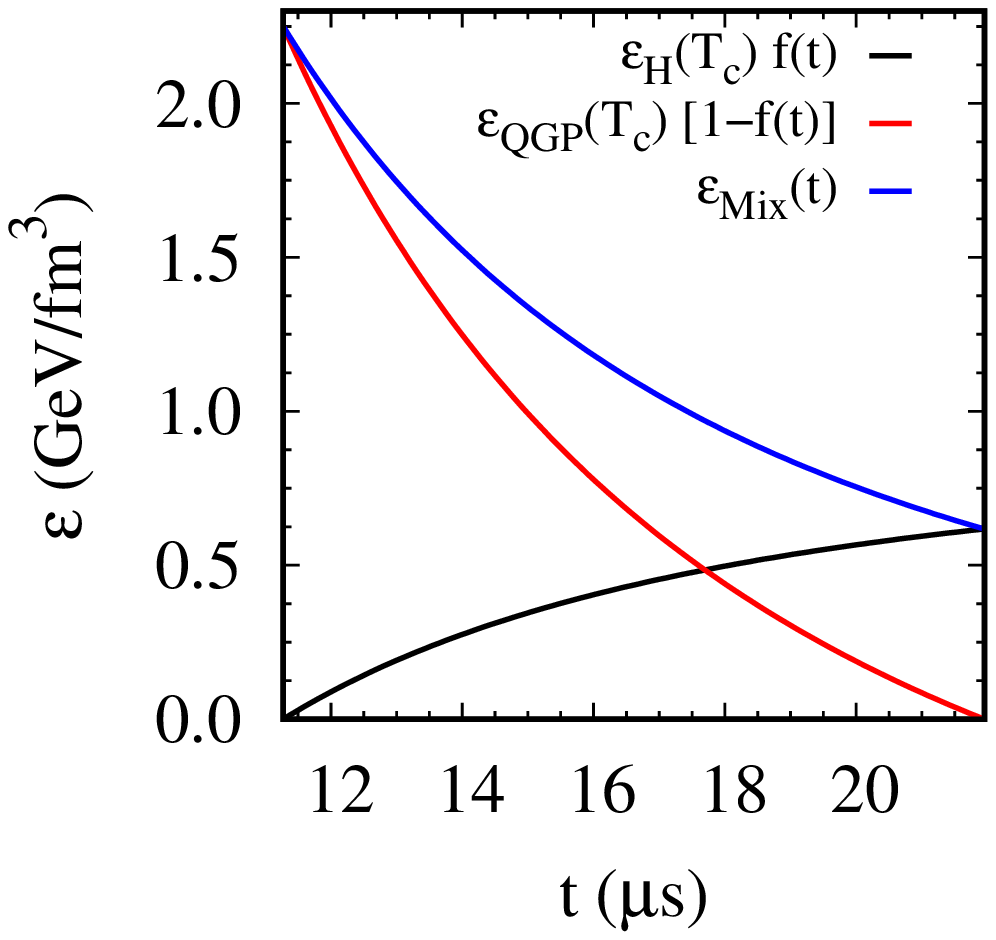}
\hspace{0.2cm}
\includegraphics[width=5cm,height=5cm]{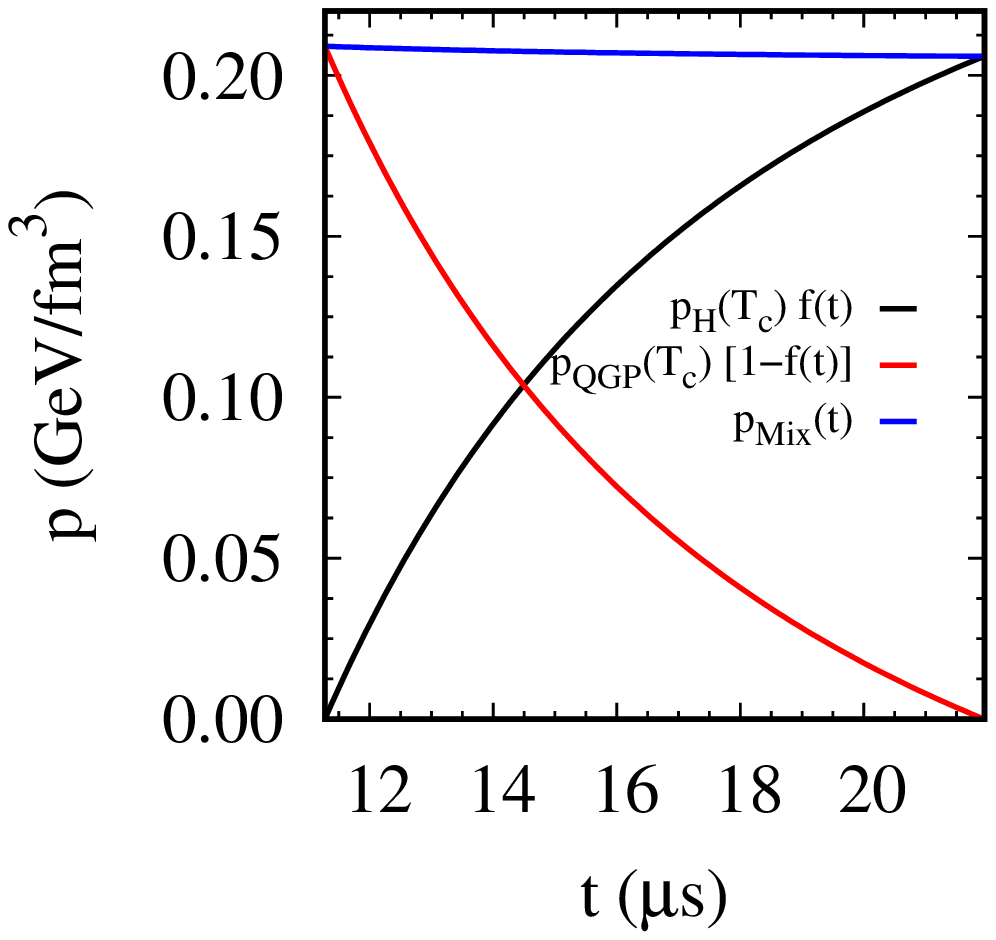}
\medskip
\caption{Time evolution of $f(t)$, the energy density and the
pressure density the mixed era of the Universe.\label{fEp}}
\end{center}
\end{figure}

In Fig(\ref{ETD}) we show the evolution of the energy density
$\varepsilon$, temperature $T$, pressure density $p$ and entropy
density $s$ with time where different colors correspond to the
different time intervals in the early eras of the universe.
Clearly, in all eras of the early Universe, these thermodynamic
parameters decrease with increasing time except at the mixed era
where temperature and pressure are constant. The temperature at
mixed era is constant and equals to $T_C$. This can be explained
as stated in Ref.\cite{Yagi:2005yb} that the release of the latent
heat recoups the cooling of the Universe due to the expansion.
Regarding the energy density and the pressure in the mixed era, we
show in Fig.(\ref{fEp}) their time evolution together with the the
function $f(t)$. Clearly, from the figure, the energy density
decreases also with time in this era while the pressure is nearly
constant with varying the time.  In the corresponding plots, the
contributions proportional to $ \varepsilon_H (T_C)$ and  $
p_H(T_C)$ increase as time runs while those proportional to $
\varepsilon_{QGP} (T_C)$ and  $ p_{QGP}(T_C)$ decrease as time
runs.  This can be attributed to the behavior of $f(t)$ seen in
the left plot in Fig.(\ref{fEp}).

\begin{figure}[t]
\begin{center}
\includegraphics[width=7cm,height=7cm]{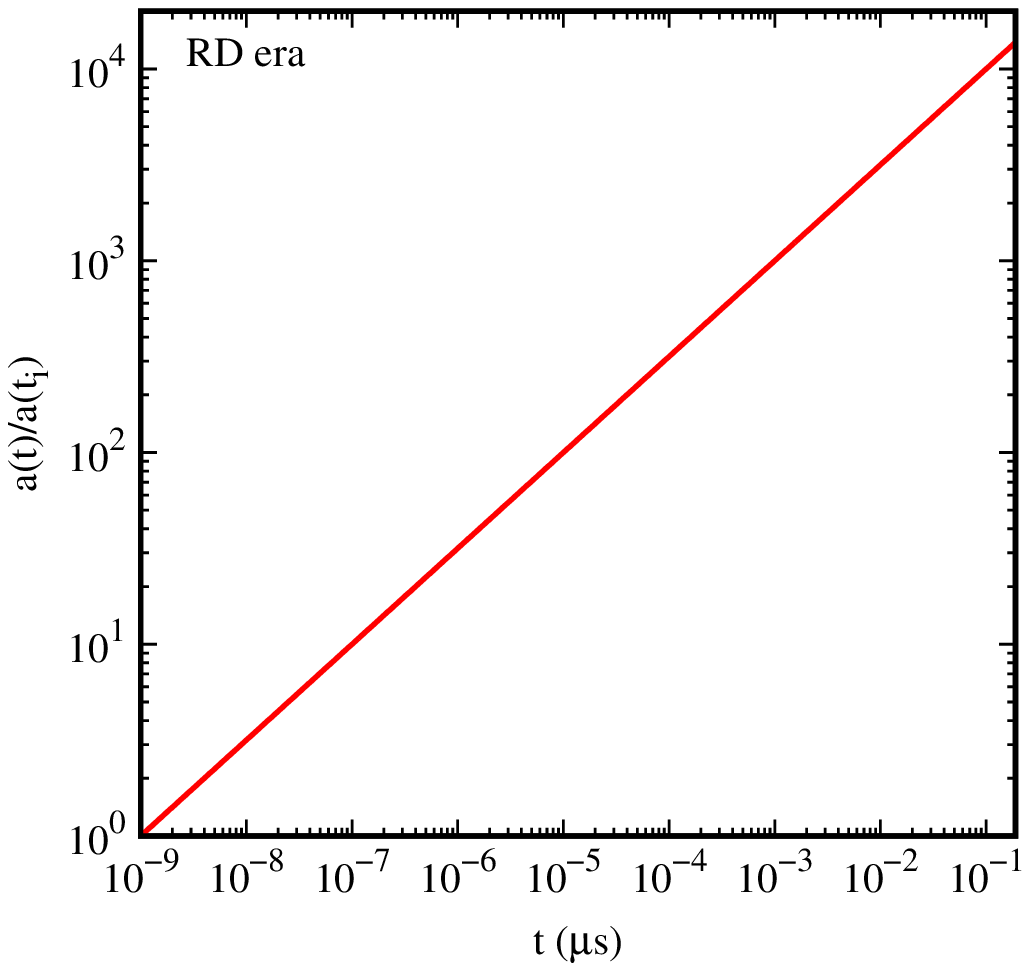}
\hspace{0.2cm}
\includegraphics[width=7cm,height=7cm]{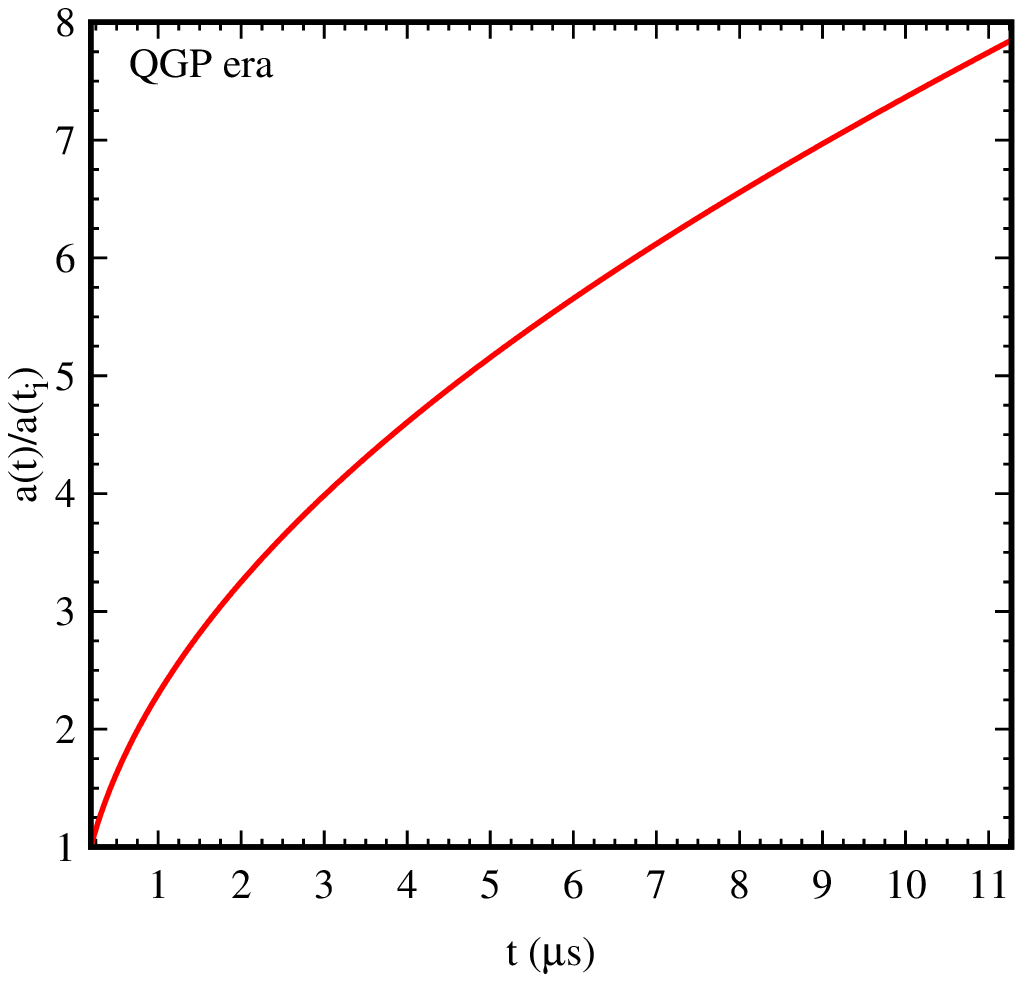}
\hspace{0.2cm}
\includegraphics[width=7cm,height=7cm]{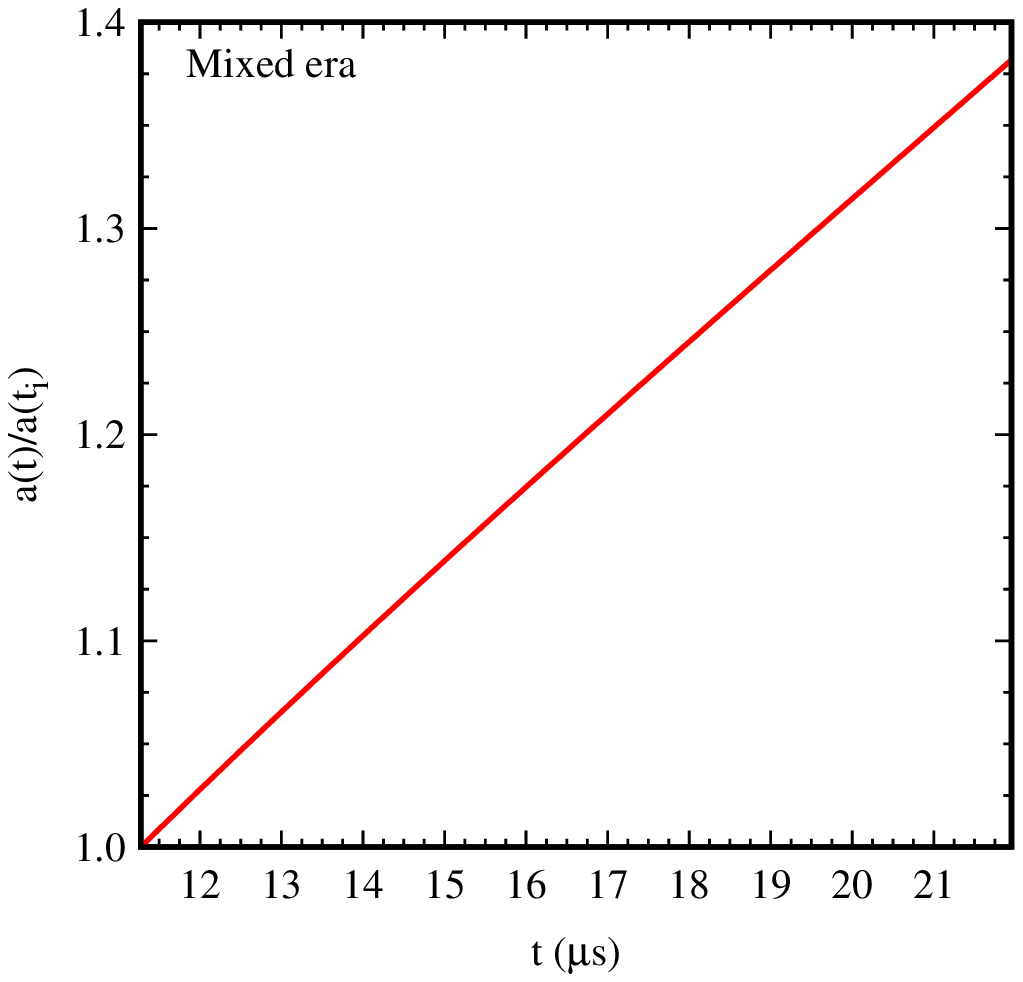}
\hspace{0.2cm}
\includegraphics[width=7cm,height=7cm]{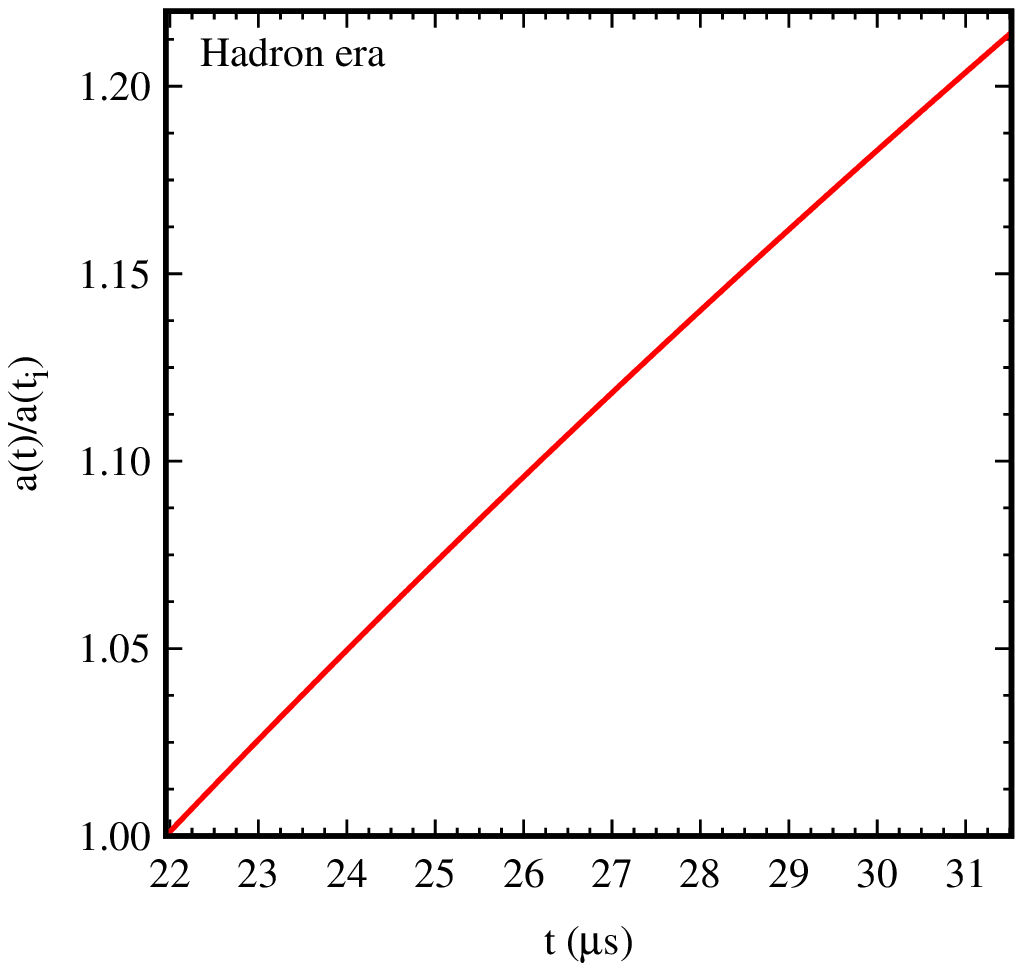}
\medskip
\caption{Time evolution of $\frac{a(t)}{a(t_i)}$ in the four eras
of the Universe where $i$ stands for the beginning time of the
given era.\label{aRDQH}}
\end{center}
\end{figure}

\begin{figure}[t]
\begin{center}
\includegraphics[width=7cm,height=7cm]{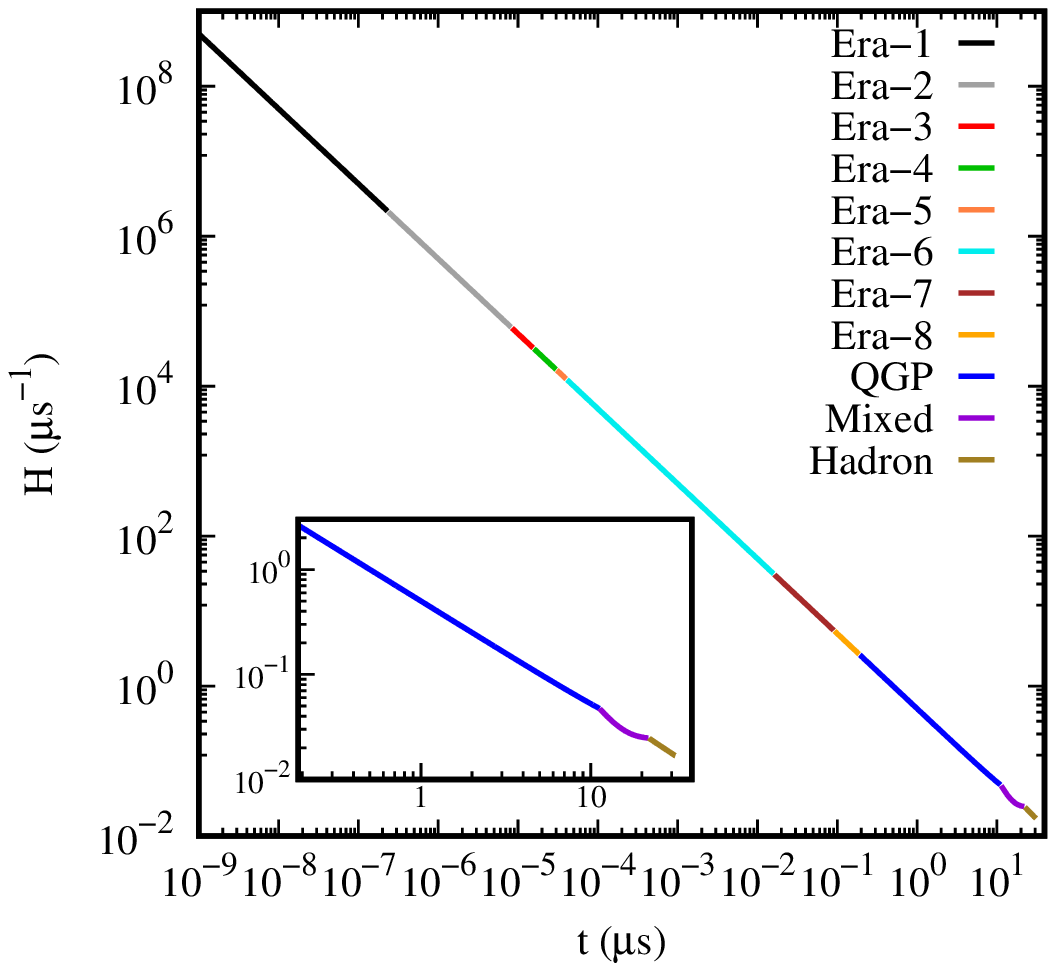}
\medskip
\caption{The time evolution of the Hubble constant in early eras
of the Universe.\label{Hub}}
\end{center}
\end{figure}

In Fig.(\ref{aRDQH}), we show the plots of the time evolution of
$\frac{a(t)}{a(t_i)}$ in the different eras of the early Universe.
Here. $a(t_i)$ stands for the scale factor at the beginning of the
concern era of the Universe. Clearly, from Fig.(\ref{aRDQH}), the
ratio $\frac{a(t)}{a(t_i)}$ increase in each era indicating
expansion of the Universe. At the end of RD, $QGP$, mixed and
hadron eras we find that $\frac{a(t_8)}{a(t_0)}\simeq 1.38\times
10^4$, $\frac{a(t_9)}{a(t_8)}\simeq7.85$,
$\frac{a(t_{10})}{a(t_9)}\simeq1.44$ and
$\frac{a(t_{11})}{a(t_{10})}\simeq1.21$.  In Fig(\ref{Hub}), we
show the time evolution of the Hubble parameter for the studied
eras of the universe where the different colors represent the
different eras. Clearly, the Hubble parameter decreases with the
increase of the time. Recall that, from Eq.(\ref{Hubp}), the
Hubble parameter is directly proportional to the square root of
the energy density. Due to the expansion of the Universe, the
volume of the Universe increases.  Since the amount of the total
energy is constant, it turns that the energy density decreases.

\section{Conclusion}
In this study we derived the analytic expressions governing the
time evolution of the thermodynamic and the cosmological
parameters in early eras of the Universe namely, the radiation
era, the quark-gluon plasma era and the hadron era. In particular,
these parameters include the energy density, the entropy density,
the temperature, the pressure in addition to Hubble parameter and
the scale factor. The values of the time corresponding to the
beginning and ending of these eras were also derived in this work.

Using the aforementioned expressions, we investigated the time
variation of the energy density, entropy density, pressure and
temperature in all these eras. Moreover, we showed the behaviors
of the Hubble parameter and scale factor with the variation of
time in the considered eras. In studying the $QGP$ era, we adopted
simple bag models for the equations of state of the thermodynamic
parameters based on the MIT bag model. However, adopting other
models with complicated equations of state of the thermodynamic
parameters can be included in our formalism even for the cases of
obtaining numerical solutions rather than analytic solutions for
the Friedmann differential equations.

\section*{Acknowledgements}
We would like to thank Ahmed Farag Ali for useful comments and
discussions.

\section*{Appendix}

\subsection{Analytic solutions for the time evolution of the
thermodynamic parameters in the $QGP$ era }

In this subsection we derive analytic solutions for the time
evolution differential equations of the energy density,
temperature and pressure.

In order to study the evolution of the energy density  with time,
in the early Universe,  we need to solve the differential equation
given in Eq.(\ref{eqdens}). To do this, firstly, we eliminate the
temperature  from  Eq.(\ref{edQGP2}) to get:
\begin{eqnarray}
p  &=& \frac{1}{3} \big (\varepsilon - 4 \mathcal{B} \big)
\label{eqs}
\end{eqnarray}
After doing the integration and making some simplifications, the
exact analytic solution of the  differential equations given in
Eq.(\ref{eqdens})  can be expressed as \bea \varepsilon_{QGP}
(t)&=& \chi(t) + \zeta(t)+\sqrt{\bigg(\chi (t)+\zeta
(t)\bigg)^2-\zeta^2 (t)} \label{edMT}\eea The functions $\zeta(t)$
and $\chi(t)$ are defined as

\bea \zeta(t) &=&- \big(1+ \eta (t)\big)\big(1-\eta
(t)\big)^{-1} \mathcal{B} \nonumber\\
\chi(t) &=& 2\bigg(1- \eta (t)\bigg)^{-2}
\mathcal{B}\label{zeta}\eea where \bea \eta(t) &=&
\exp\bigg(4\,\sqrt{\frac{8 \pi \mathcal{B}\, G}{3}}\,\, t + \xi
\bigg) \label{lamd}\eea  the quantity $\xi$ is defined through
\bea \xi &=& \ln\bigg(\frac{\sqrt{\varepsilon_{RD}
(t_8)}+\sqrt{\mathcal{B}}}{\sqrt{\varepsilon_{RD}
(t_8)}-\sqrt{\mathcal{B}}}\bigg)-4\,\sqrt{\frac{8 \pi
\mathcal{B}\, G}{3}}\,\, t_8 \label{xi}\eea
where $\varepsilon_{RD}(t_8)$ will be equivalent to the initial
value of the energy density at the time $t_8$ of the beginning of
the QGP era.

We turn now to derive analytic solution of the temperature for the
bag model of QGP discussed above. Firstly,  the differential
equation governs the time evolution of the temperature, after
eliminating the pressure and energy density using the equations of
the state in Eq.(\ref{edQGP2}) and upon substituting in
Eq.(\ref{eqdens}), can be written as

\begin{equation}
\frac{dT}{T%
\sqrt{T^{4}+\frac{\mathcal{B}}{N_{QGP}\frac{\pi^2}{30}}}}= -
\frac{2}{3} \sqrt{\,N_{QGP}\frac{\pi^3}{5} G}\,\,dt, \label{eq05}
\end{equation}%
We can integrate Eq.(\ref{eq05}) to get

\begin{equation}
\int_{T_{8}}^{T}\frac{dT}{T
\sqrt{T^{4}+\frac{\mathcal{B}}{N_{QGP}\frac{\pi^2}{30}}}}= - \frac{2}{3} \sqrt{\,N_{QGP}\frac{\pi^3}{5} G}\,\left( t-t_{8}\right),
\label{eq055}
\end{equation}%
with $t_8\simeq 1.85\times 10^{-7}\,s$ and $T_8=T_{RD}(t_8)$ are
the initial time and initial temperature at the beginning of the
QGP phase respectively. We find that the differential equation
above has a solution in the form
\begin{equation}
\ln\bigg(\frac{T^2}{\frac{\mathcal{B}}{\frac{\pi^2}{30}N_{QGP}}+\sqrt{\frac{\mathcal{B}}{\frac{\pi^2}{30}N_{QGP}}
T^4+\frac{\mathcal{B}^2}{\frac{\pi^4}{900}N^2_{QGP}}}}\bigg)-
\ln\bigg(\frac{T^2_{8}}{\frac{\mathcal{B}}{\frac{\pi^2}{30}N_{QGP}}+\sqrt{\frac{\mathcal{B}}{\frac{\pi^2}{30}N_{QGP}}
T^4_8+\frac{\mathcal{B}^2}{\frac{\pi^4}{900}N^2_{QGP}}}}\bigg) = -
\frac{4}{3} \sqrt{6\pi\, B G}\,\left( t-t_{8}\right) \label{eq06}
\end{equation}
which can be expressed as

\begin{equation} T_{QGP}(t)= \sqrt{\frac{2\, \mathcal{B}\kappa(t)}{\big(\frac{\pi^2}{30}N_{QGP}-\mathcal{B}\, \kappa^2 (t)
\big)}}\label{Tem1}\end{equation} where the function $\kappa(t)$
is given as
\begin{equation} \kappa(t) = b \exp\big[{-\frac{4}{3} \sqrt{6\pi \mathcal{B} G}\left( t-t_{i}\right)}\big]
\end{equation}
with \be b=
T^2_8\bigg(\frac{\mathcal{B}}{\frac{\pi^2}{30}N_{QGP}}+\sqrt{\frac{\mathcal{B}}{\frac{\pi^2}{30}N_{QGP}}
T^4_8+\frac{\mathcal{B}^2}{\frac{\pi^4}{900}N^2_{QGP}}}\bigg)^{-1}\ee

The time evolution of the pressure can be derived using
Eqs.(\ref{eqs}) as we show in the following. Recall that from
Eqs.(\ref{eqs}) we have

\begin{equation}
p = \frac{1}{3} \big (\varepsilon - 4 \mathcal{B} \big)
\label{eqss}
\end{equation}
so we get
\begin{equation}
\varepsilon =3 p +4 \mathcal{B}  \label{eqsss}
\end{equation}
after substituting in the differential equation given in
Eq.(\ref{eqdens}) we get
\begin{equation}
\frac{d \,p}{\left( p + \mathcal{B} \right)\sqrt{3 \big( p +
\mathcal{B}\big)+ \mathcal{B} }} =- 4\, \sqrt{\frac{8 \pi G}{3}}
dt \label{eq444}
\end{equation}
defining $y= p + \mathcal{B}$ we obtain
\begin{equation}
\frac{d y}{y \sqrt{3 y+ \mathcal{B} }} =- 4\, \sqrt{\frac{8 \pi
G}{3}} dt \label{eq666}
\end{equation}
After performing the integration we find that
\begin{equation}
\bigg\{\ln\bigg[\frac{\sqrt{\mathcal{B}}-\sqrt{3 p +4 \mathcal{B}
}}{\sqrt{\mathcal{B}}+\sqrt{3 p +4 \mathcal{B}
}}\bigg]\bigg\}^{p}_{p_8} =- 4\, \sqrt{\frac{8 \pi \mathcal{B}
\,G}{3}} \big( t-t_{8}\big) \label{eq77}
\end{equation}
which can be written as
\begin{equation}
\frac{\sqrt{\mathcal{B}}-\sqrt{3 p +4 \mathcal{B}
}}{\sqrt{\mathcal{B}}+\sqrt{3 p +4 \mathcal{B} }}
=\frac{\sqrt{\mathcal{B}}-\sqrt{3 p_{8} +4 \mathcal{B}
}}{\sqrt{\mathcal{B}}+\sqrt{3 p_{8} +4 \mathcal{B} }}\,\exp\bigg[-
4\, \sqrt{\frac{8 \pi \mathcal{B} \, G}{3}} \big(
t-t_{8}\big)\bigg] \label{eq555}
\end{equation}
defining $\varrho_{8}= \frac{\sqrt{\mathcal{B}}-\sqrt{3 p_{8} +4
\mathcal{B} }}{\sqrt{\mathcal{B}}+\sqrt{3 p_{8} +4 \mathcal{B} }}$
and $k=\exp\bigg[- 4\, \sqrt{\frac{8 \pi \mathcal{B} \, G}{3}}
\big( t-t_{8}\big)\bigg]$ and solve for $p$ we get

\begin{equation}  p= \frac{-3 \varrho_{8}^2 \mathcal{B} k^2-10 \varrho_{8} \mathcal{B} k-3 \mathcal{B}}
{3 \big(\varrho_{8} k+1\big)^2}\end{equation} The explicit
dependency of the pressure on the time is clear as, after
substituting $k$, $p$ have finally the form
\begin{equation}  p= \frac{-3 \varrho_{8}^2 \mathcal{B} \exp\bigg[- 8\, \sqrt{\frac{8 \pi \mathcal{B} \, G}{3}} \big(
t-t_{8}\big)\bigg] -10 \varrho_{8} \mathcal{B} \exp\bigg[- 4\,
\sqrt{\frac{8 \pi \mathcal{B} \, G}{3}} \big( t-t_{8}\big)\bigg]
-3 \mathcal{B}} {3 \bigg(\varrho_{8} \exp\big[- 4\, \sqrt{\frac{8
\pi \mathcal{B} \, G}{3}} \big( t-t_{8}\big)\big]
+1\bigg)^2}\end{equation}

\subsection{Derivation of some relations in the Mixed era}

As discussed in subsection \ref{mixsub}, the energy density and
the pressure in the mixed phase can be parameterized as \bea
\varepsilon_{mix}(t) &=& \varepsilon_{H}(T_C) f(t) +
\varepsilon_{QGP} (T_C)\big(1-f(t) \big)\nonumber\\ p_{mix}(t) &=&
p_{H}(T_C) f(t) + p_{QGP} (T_C)\big(1-f(t) \big) \label{qgmix1}
\eea In terms of $\mathcal{B}$ and $r$ the the energy density and the
pressure in the above equations are given as:
\begin{eqnarray}
\varepsilon_{mix} &=& \mathcal{B} \bigg( 1 - 4f + \frac{3r}{r-1} \bigg) \nonumber\\
p_{mix} &=& \mathcal{B} \bigg( \frac{r}{r-1} - 1 \bigg) \nonumber\\
\frac{d\varepsilon_{mix}}{dt} &=& -4\mathcal{B}\frac{df}{dt}
\label{eqftEquationSet1}
\end{eqnarray}
additionally, from Eq.(\ref{eqdens}), we can write
\begin{equation}
-\frac{d\varepsilon_{mix}/dt}{3 \sqrt{\varepsilon_{mix}} \left(
\varepsilon_{mix} + p_{mix}\right)} = \frac{1}{\lambda \sqrt{\mathcal{B}}}
\label{eqdensMix1}
\end{equation}
Substituting the components of Eq.(\ref{eqftEquationSet1}) into
Eq.(\ref{eqdensMix1}) we obtain the below differential equation
for $f(t)$
\begin{equation}
    \frac{df}{dt} = \frac{3}{\lambda} \left( \frac{r}{r-1} - f\right)\, \sqrt{4 (1-f) + \frac{3}{r-1}}
\end{equation}

\end{document}